\documentclass[preprint2]{proto}
\bibpunct{(}{)}{;}{a}{,}{,}
\usepackage{natbib}
\usepackage{times}

\usepackage{color}
\usepackage{amsbsy}
\usepackage{amsmath}

\begin{document}

\title{\textbf{\LARGE New Paradigms For Asteroid Formation}}

\author {\textbf{\large Anders Johansen}}
\affil{\small\em Lund University}

\author {\textbf{\large Emmanuel Jacquet}}
\affil{\small\em Canadian Institute for Theoretical Astrophysics, University of
Toronto}

\author {\textbf{\large Jeffrey N.\ Cuzzi}}
\affil{\small\em NASA Ames Research Center}

\author {\textbf{\large Alessandro Morbidelli}}
\affil{\small\em University of Nice-Sophia Antipolis, CNRS}

\author {\textbf{\large Matthieu Gounelle}}
\affil{\small\em Mus\'eum National d'Histoire Naturelle, Institut Universitaire
de France}

\begin{abstract}
\baselineskip = 11pt
\leftskip = 0.65in 
\rightskip = 0.65in
\parindent=1pc
{\small Asteroids and meteorites provide key evidence on the formation of
planetesimals in the Solar System. Asteroids are traditionally thought to form
in a bottom-up process by coagulation within a population of initially km-scale
planetesimals. However, new models challenge this idea by demonstrating that
asteroids of sizes from 100 to 1000 km can form directly from the gravitational
collapse of small particles which have organised themselves in dense filaments
and clusters in the turbulent gas. Particles concentrate passively between
eddies down to the smallest scales of the turbulent gas flow and inside
large-scale pressure bumps and vortices. The streaming instability causes
particles to take an active role in the concentration, by piling up in dense
filaments whose friction on the gas reduces the radial drift compared to that
of isolated particles. In this chapter we review new paradigms for asteroid
formation and compare critically against the observed properties of asteroids
as well as constraints from meteorites. Chondrules of typical sizes from 0.1 to
1 mm are ubiquitous in primitive meteorites and likely represent the primary
building blocks of asteroids. Chondrule-sized particles are nevertheless
tightly coupled to the gas via friction and are therefore hard to concentrate
in large amounts in the turbulent gas. We review recent progress on
understanding the incorporation of chondrules into the asteroids, including
layered accretion models where chondrules are accreted onto asteroids over
millions of years. We highlight in the end ten unsolved questions in asteroid
formation where we expect that progress will be made over the next decade.
 \\~\\~\\~}

\end{abstract}  

\section{\textbf{INTRODUCTION}}

The Solar System contains large populations of pristine planetesimals that have
remained relatively unchanged since their formation. Our proximity to the
asteroid belt provides astronomers, planetary scientists  and cosmochemists
access to extremely detailed data about asteroid compositions, sizes and
dynamics. Planetesimals are the building blocks of both terrestrial planets
and the cores of the giant planets, as well as the super-Earths (with various
degrees of gaseous envelopes) which are now known to orbit around a high
fraction of solar-type stars \citep{Fressin+etal2013}. The formation of
planetesimals is thus a key step towards the assembly of planetary systems, but
many aspects of the planetesimal formation process remain obscure.

Recent progress in understanding planetesimal formation was triggered by two
important realisations. The first is that macroscopic dust particles (mm or
larger) have poor sticking properties. Laboratory experiments and coagulation
models show that it is difficult to form planetesimals by direct sticking of
silicate particles, most importantly because particle growth stalls at
millimeter sizes where the particles bounce off each other rather than stick
\citep{Guettler+etal2010}. While some particle growth is still possible at
relatively high collision speeds, due to the net transfer of mass from small
impactors onto large targets \citep{Wurm+etal2005,Windmark+etal2012a}, the
resulting growth rate is too low to compete with the radial drift of the
particles.

The second important realisation is that particles are concentrated to very
high densities in the turbulent gas flow. This idea is not new --
\cite{Whipple1972} already proposed that large-scale pressure bumps can trap
particles, as their radial drift speed vanishes in the pressure bump where the
radial pressure gradient is zero. However, the advent of supercomputing led to
the discovery and exploration of a large number of particle concentration
mechanisms. Large-scale axisymmetric pressure bumps, akin to those envisioned
by \cite{Whipple1972}, have been shown to arise spontaneously in simulations of
protoplanetary disk turbulence driven by the magnetorotational instability
\citep{Johansen+etal2009a,Simon+etal2012}. Particle densities reach high enough
values inside these pressure bumps to trigger gravitational collapse to form
planetesimals with sizes up to several 1000 km
\citep{Johansen+etal2007,Johansen+etal2011,Kato+etal2012}. The baroclinic
instability, which operates in the absence of coupling between gas and magnetic
field, leads to the formation of slowly overturning large-scale vortices
\citep{KlahrBodenheimer2003} which can act as dust traps in a similar way as
pressure bumps \citep{BargeSommeria1995}.

In the streaming instability scenario the particles play an active role in the
concentration \citep{YoudinGoodman2005}. The relative motion between gas and
particles is subject to a linear instability whereby axisymmetric filaments of
a slightly increased particle density accelerate the gas towards the Keplerian
speed and hence experience reduced radial drift. This leads to a run-away pile
up of fast-drifting, isolated particles in these filaments
\citep{JohansenYoudin2007}. The densities achieved can be as high as 10,000
times the local gas density \citep{BaiStone2010b,Johansen+etal2012}, leading to
the formation of planetesimals with characteristic diameters of 100-200 km for
particle column densities relevant for the asteroid belt, on a time-scale of
just a few local orbital periods.

A concern about large-scale particle concentration models is that typically
very large particles are needed for optimal concentration (at least dm in size
when the models are applied to the asteroid belt). Chondrules of typical sizes
from 0.1 to 1 mm are ubiquitous in primitive meteorites, but such small
particles are very hard to concentrate in vortices and pressure bumps or
through the streaming instability. One line of particle concentration models
has nevertheless been successful in concentrating chondrules. Swiftly rotating
low pressure vortex tubes expel particles with short friction times
\citep{SquiresEaton1990,SquiresEaton1991,Wang_Maxey_1993}. This was proposed to
explain the characteristic sizes and narrow size ranges of chondrules observed
in different chondrites \citep{Cuzzi+etal2001} and lead to the formation of
100-km-scale asteroids from rare high-density concentrations
\citep{Cuzzi+etal2008,Cuzzi+etal2010}. However, the evaluation of the
probability for such high-density concentrations to occur over sufficiently
large scales depends on scaling computer simulations to the very large
separations between the energy injection scale and the dissipation scale
relevant for protoplanetary disks; \cite{Pan+etal2011} found that the particle
clustering gets less contribution from the addition of consecutively larger
scales than originally thought in the model of
\cite{Cuzzi+etal2008,Cuzzi+etal2010}.

Therefore the incorporation of chondrules into the asteroids is still an
unsolved problem in asteroid formation. This is one of the main motivations for
this review. We refer the readers to several other recent reviews on the
formation of planetesimals which provide a broader scope of the topic beyond
asteroid formation
\citep[e.g.][]{CuzziWeidenschilling2006,ChiangYoudin2010,Johansen+etal2014}. 

The review is organised as follows. The first two sections discuss the
constraints on asteroid formation from the study of meteorites (Section
\ref{s:meteorites}) and asteroids (Section \ref{s:asteroids}). In Section
\ref{s:dust_growth} we review laboratory experiments and computer simulations
of dust coagulation to illustrate the formidable barriers which exist to
planetesimal formation by direct sticking. The turbulent concentration model,
in which chondrule-sized particles are concentrated at the smallest scales of
the turbulent gas flow, is discussed in Section
\ref{s:turbulent_concentration}. The following Section \ref{s:streaming} is
devoted to particle concentration in large-scale pressure bumps and through
streaming instabilities. In Section \ref{s:layered} we discuss layered
accretion models where the chondrules are accreted onto planetesimals over
millions of years. Finally, in Section \ref{s:questions} we pose ten open
questions in the formation of asteroids on which we expect major progress in
the next decade.

\section{\textbf{CONSTRAINTS FROM METEORITES}}
\label{s:meteorites}

Meteorites provide a direct view of the solid material from which the asteroids
accumulated, while the crystallisation ages of the component particles and the
degree of heating and differentiation of the parent bodies give important
information about the time-scales for planetesimal formation in the Solar
System.

Meteorites may be broadly classified in two categories
\citep{Weisbergetal2006}: primitive meteorites (also known as chondrites) and
differentiated meteorites. \textit{Chondrites}, which make up 85\% of the
observed falls, are basically collections of mm- and sub-mm-sized solids,
little modified since agglomeration and lithification (compression) in their
parent bodies. They exhibit nonvolatile element abundances comparable to the
solar photosphere's \citep{PalmeJones2005}. \textit{Differentiated meteorites}
derive from parent bodies which underwent significant chemical fractionations
on the scale of the parent body, resulting in the asteroid-wide segregation of
an iron core and silicate mantle and crust. In the process, differentiated
meteorites have lost not only their accretionary (presumably chondritic)
texture, but also their primitive chemical composition, for, depending on which
part of the parent body they sample, some may be essentially pure metal (the
\textit{iron meteorites}) while others are essentially metal-free (the
\textit{achondrites}).

It is among the components of chondrites that the oldest solids of the solar
system, the \textit{refractory inclusions} \citep{Krotetal2004,MacPherson2005},
in particular Calcium-Aluminium-rich Inclusions (CAI), have been identified.
Their age of 4567.3$\pm$0.16 Myr \citep{Connelly+etal2012} marks the commonly
accepted ``time zero'' of the Solar System. But the ubiquitous components of
chondrites are the eponymous \textit{chondrules}
\citep{Hewinsetal2005,ConnollyDesch2004}, which are millimeter-size silicate
spherules presumably resulting from transient (and repeated) high-temperature
episodes in the disk, but whose very nature remains a long-standing
cosmochemical and astrophysical enigma \citep{Boss1996,Desch+etal2012}. All
these components are set in a fine-grained matrix comprised of amorphous and
crystalline grains native to the disk as well as rare presolar grains
\citep{Brearley1996}. While compositionally primitive, chondrites may have
undergone some degree of thermal metamorphism, aqueous alteration and shock
processing on their parent body.
\begin{figure}[!t]
  \includegraphics[width=\linewidth]{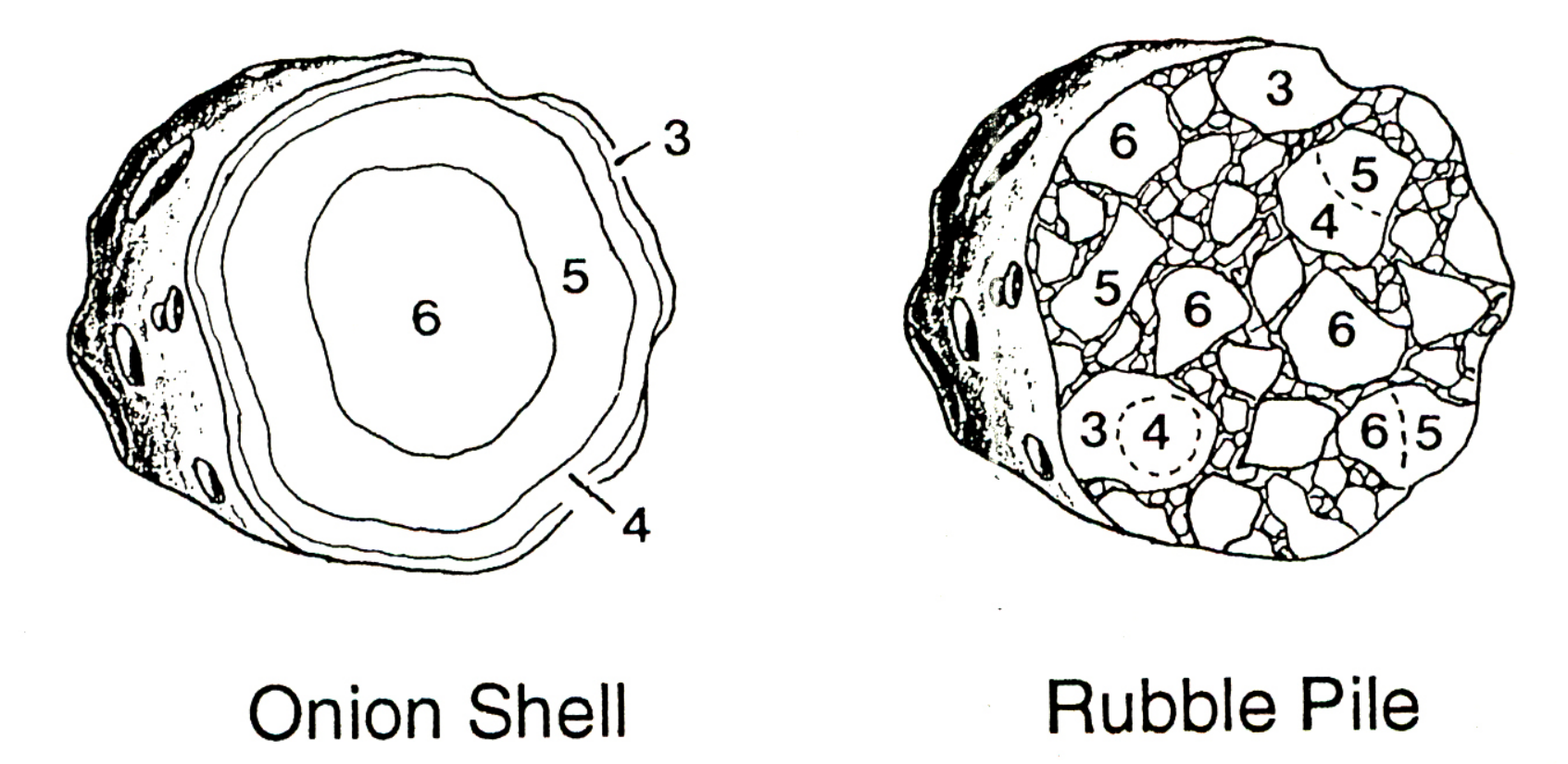}
  \caption{The left panel indicates the initial state of a parent body of 100
  km in radius, which was initially homogeneous throughout, after heating by
  $^{26}$Al. The proportions of the least (3) to most (6) altered material are
  constrained by meteorite statistics. The right panel indicates the same body
  after catastrophic fragmentation and reassembly, as a rubble pile with some
  highly altered material now near the surface. This body is then cratered by
  subsequent impacts, releasing samples of all metamorphic grades
  \citep[adapted from][]{Scott_Rajan_1981}.}
  \label{f:onion_rubble}
\end{figure}

Despite their general petrographic similarity and roughly solar composition,
chondrites are actually quite variable, and 14 distinct \textit{chemical
groups} have been recognized so far, each of which believed to represent a
single parent body -- sometimes with supporting evidence from cosmic-ray
exposure ages \citep{Eugster+etal2006} -- or at least a family of parent bodies
formed in the same nebular \textit{reservoir} (i.e.\ a compositionally
distinctive space-time section of the disk). There are various levels of
affinities between these groups (e.g.\ clans, classes) but we will be here
content to distinguish \textit{carbonaceous chondrites} [henceforth ``CCs'';
comprising the CI, CM, CO, CV, CK, CR, CH, CB groups] from
\textit{non-carbonaceous chondrites} [henceforth ``EORs''; comprising the
enstatite (EH, EL), the ordinary (H, L, LL) and the Rumuruti (R) chondrite
groups]. Carbonaceous chondrites are more ‘‘primitive’’ in the sense that they
have a solar Mg/Si ratio and a $^{16}$O-rich oxygen isotopic composition closer
to that of the Sun \citep[e.g.][]{ScottKrot2003}. Non-carbonaceous chondrites,
though poorer in refractory elements, are more depleted in volatile elements,
have subsolar Mg/Si ratios and a more terrestrial isotopic composition for many
elements \citep{Trinquieretal2009}. EORs have generally undergone thermal
metamorphism (see sketch in Figure \ref{f:onion_rubble}) while aqueous
alteration has been prevalent in CCs \citep{Hussetal2006,Brearley2003}, but
there are again exceptions.

\subsection{Primary texture and aerodynamic sorting}

The texture of most chondrites has been reworked by impact fragmentation and
erosion on their parent body. However, rare pieces of CM and CO chondrites have
been found, referred to as ``primary texture''
\citep{Metzler_etal_1992,Brearley1993}, which seem to retain the nature of a
pre-brecciated body. Primary texture appears to consist of nothing but
dust-rimmed chondrules of very similar properties, loosely pressed together.

The constituents of most chondrites appear well-sorted by size, with strong
mean size differences from one group to another \citep{BrearleyJones1998}.
Whether these differences arise from some regionally or temporally variable
bouncing-barrier \citep{Jacquet2014size}, some aerodynamic sorting process
(Sections 5 and 7), or some aspect of the mysterious chondrule formation
process itself, they provide an important clue to primary accretion.
\cite{Hezeletal2008} have emphasized the need for better particle counting
statistics, and indeed one recent chondrule size distribution measurement taken
from Allende, of a far larger sample than analyzed previously
\citep{Fisher_etal_2014}, points to a distribution substantially broader for
that chondrite than previously reported.

Aerodynamic sorting has been suggested often as an important factor in
selecting for the contents of primary texture \citep[see][for a
review]{CuzziWeidenschilling2006}. Comparing the aerodynamical friction time of
objects of greatly different density, such as silicate and metal grains, shows
that their friction times are quite similar in the least altered meteorites
\citep{Dodd1976,Kuebler_etal_1999}, suggesting that asteroids selectively
incorporated components with specific aerodynamical properties (we discuss this
further in Section \ref{s:homo}).

\subsection{\textbf{The abundance and distribution of $^{26}$Al and $^{60}$Fe}}
\label{s:SLR}

The melting of the parent bodies of differentiated meteorites puts important
constraints on the time-scale for planetesimal formation in the asteroid belt.
While electromagnetic heating \citep{Sonett1968} or impact heating
\citep{Keil1997} have been considered in the literature, the most likely source
of planetesimal heating is the decay of the short-lived radionuclides (SLRs)
$^{26}$Al (with mean life-time $\tau$ = 1.0 Myr) and $^{60}$Fe ($\tau$ = 3.7
Myr) \citep{Urey1955}. Depending on their respective initial abundance, and on
the time of planetesimal accretion, both could have significantly contributed
to planetesimal heating. Additionally, short-lived nuclides provide
crystallisation ages which can be calibrated using a long-lived radionuclide
decay system such as Pb-Pb, under the assumption that the short-lived
radionuclide was homogeneously distributed in the solar protoplanetary disk.

The content of SLRs in CAIs is usually identified to that of the nascent Solar
System \citep{Dauphas2011}. Excesses of $^{26}$Mg linearly correlating with
$^{27}$Al content were first observed in an Allende CAI in 1976
\citep{Lee1976}. This isochron diagram demonstrated the presence of $^{26}$Al
in the nascent Solar System. The CAIs from a diversity of chondrite groups
formed with an initial ($^{26}$Al/$^{27}$Al)$_0$ of roughly 5 $\times$
10$^{-5}$ \citep{MacPherson2014}. A remarkably tight isochron for CAIs in the
CV chondrites was obtained by \cite{Jacobsen2008}. The deduced
($^{26}$Al/$^{27}$Al)$_0$  ratio of (5.23$\pm$0.13)$\times$10$^{-5}$ is often
considered as the initial value for the Solar System and the small dispersion
as indicative of a narrow formation interval ($\leq$ 40,000 yr). However, this
interpretation should be limited to the region where the unusually large CV
CAIs have formed \citep{Krot2009}. This is especially true since many CAIs are
known to have formed without any $^{26}$Al \citep{Liu2012hib,Makide2013}. This
indicates some level of heterogeneity in the $^{26}$Al distribution within the
region where CAIs formed (assuming that region was unique, which is supported
by the ubiquitous $^{16}$O enrichment of CAIs compared to e.g.\ chondrules).
\cite{Larsen_etal_2011} used bulk magnesium isotopic measurements to suggest
that the heterogeneity of $^{26}$Al distribution might have reached 80\% of the
canonical value in the solar protoplanetary disk. However, \cite{Kitaetal2013}
and \cite{Wasserburg_etal_2012} argue that the observed variations can be
better ascribed to small heterogeneities in the stable isotope $^{26}$Mg.

Although the presence of live $^{60}$Fe in the early Solar System was
demonstrated almost 20 years ago \citep{Shuko1993}, the determination of the
Solar System initial abundance is complicated by the difficulty of obtaining
good isochrons for CAIs \citep{Quitte2007}, given their low abundance in Ni. To
bypass that difficulty most measurements were performed on chondrules which are
believed to have formed from around the same time as CAIs up to three million
years later \citep{Connelly+etal2012}. High initial ($^{60}$Fe/$^{56}$Fe)$_0$
ratios were originally reported in chondrules from unequilibrated ordinary
chondrites (UOC) which experienced very little heating or aqueous alteration
\citep[e.g.,][]{Tachibana2003, Mostefaoui2005, Tachibana2006}.
\citet{Telus2012} showed that most of these previous data were statistically
biased and that most chondrules do not show any $^{60}$Ni excesses indicative
of the decay of $^{60}$Fe. The high values obtained in older publications could
be due to statistical biases related to low counts \citep{Telus2012} or to
thermal metamorphism which would have lead to the redistribution of Ni isotopes
\citep{Chaussidon2009}. Recently improved techniques for measuring bulk
chondrules in unequilibrated ordinary chondrites have yielded initial Solar
System values for ($^{60}$Fe/$^{56}$Fe)$_0$ of $\approx$1$\times$10$^{-8}$
\citep{Tang2012, Chen2013}. This value is consistent with that inferred from
Fe-Ni isotope measurements of a diversity of differentiated meteorites
\citep{Quitte2011, Tang2012}.

In conclusion, it seems that most CAIs formed with an initial ratio
($^{26}$Al/$^{27}$Al)$_0$ of roughly 5 $\times$ 10$^{-5}$, which can be
considered in a first approximation as the Solar System average or typical
initial value. Some heterogeneity was undoubtedly present, but its exact level
is still unknown. On the other hand, it is likely that the initial
($^{60}$Fe/$^{56}$Fe)$_0$ of the Solar System was lower than
1$\times$10$^{-8}$, though high levels of $^{60}$Fe (up to $10^{-6}$) have been
detected in some components of chondrites. At the time of writing this review
there was no evidence of $^{60}$Fe heterogeneity within the solar
protoplanetary disk.

\subsection{\textbf{The origin of $^{26}$Al and $^{60}$Fe}}

The initial Solar System ratio ($^{26}$Al/$^{27}$Al)$_0$ of roughly 5
$\times$10$^{-5}$ is well above the calculated average Galactic abundance
\citep{Huss2009}. This elevated abundance indicates a last minute origin for
$^{26}$Al. Production of $^{26}$Al by irradiation has been envisioned in
different contexts \citep[e.g.,][]{Lee1978,Gounelle2006}, but fails to produce
enough $^{26}$Al relative to $^{10}$Be ($\tau$ = 2.0 Myr), another SLR whose
origin by irradiation is strongly supported by experimental data
\citep{Gounelle2013}. This leaves stellar delivery as the only possibility for
$^{26}$Al introduction in the Solar System. Though Asymptotic Giant Branch
Stars produce elevated amounts of $^{26}$Al \citep{Lugaro2012}, it is extremely
unlikely that stars at this evolutionary stage are present in a star forming
region \citep{Kastner1994}. Thus massive stars are the best (unique) candidates
for the origin of the Solar System's $^{26}$Al.

The recently obtained lower estimates for the abundance of $^{60}$Fe (see
Section \ref{s:SLR}) are compatible with a Galactic background origin,
independently of whether this is calculated crudely using a box model
\citep{Huss2009}, or taking into account the stochastic nature of star
formation in molecular clouds \citep{Gounelle2009}. However, if the initial
$^{60}$Fe abundance corresponds instead to a much higher
($^{60}$Fe/$^{56}$Fe)$_0$ ratio of $\approx$ 10$^{-6}$, a last minute origin is
needed. Irradiation processes cannot account for $^{60}$Fe production, because
of its richness in neutrons \citep{Lee1998}. The winds of massive stars can
also be excluded, given their low abundance in $^{60}$Fe. In contrast,
supernovae are copious producers of $^{60}$Fe, essentially because this SLR is
synthesized in abundance during the hydrostatic and explosive phases of such
massive stars \citep{Woosley2007}. Two astrophysical settings have been
envisioned so far. In the first (classical) model \citep{Cameron1977}, the
supernova ejecta hits a molecular cloud core and provokes its gravitational
collapse \citep{Boss2013} as well as injecting $^{60}$Fe and other SLRs. A
newer model injects $^{60}$Fe in an already formed protoplanetary disk
\citep{Hester2004,Ouellette2007}. In either cases, the supernova progenitor
mass is in the range 20-60 M$_{\sun}$, because too massive supernovae are
extremely rare and very disruptive to their environment \citep{Chevalier2000}.
Alternatively, generations of supernovae could have enriched the gas of the
giant molecular cloud, so that the solar protoplanetary disk simply inherited
the elevated abundances of the birth cloud \citep{Vasileiadis+etal2013}.

All these models suffer from an important difficulty, namely the overabundance
of $^{60}$Fe in supernova ejecta relative to $^{26}$Al. The Solar System
$^{60}$Fe/$^{26}$Al mass ratio was either 4 $\times$ 10$^{-3}$ or 0.4 depending
on the adopted initial $^{60}$Fe abundance. In any case this is well below the
$^{60}$Fe/$^{26}$Al mass ratio predicted by supernovae nucleosynthetic models
whose variation domain extends from 1.5 to 5.5 for a progenitor mass varying
between 20 and 60 M$_{\sun}$. Heterogeneity in the composition of supernova
ejecta has been proposed as a possible solution to that discrepancy
\citep{Pan2012}. However that variability is limited to a factor of 4 which
would help resolve the discrepancy only marginally in the case of the
(unlikely) high $^{60}$Fe value. Finally, the astrophysical context of any
supernova model is difficult to reconcile with observations of star-forming
regions \citep{Williams2007,GounelleMeibom2008}. The commonly proposed setting
for supernova contamination of a protoplanetary disk or or a dense core is
similar to that of the Orion Nebula where disks are seen within a few tenths of
parsec of the massive star $\theta ^1$ C Ori. The problem with that setting is
that, when $\theta ^1$ C Ori will explode in 4 Myr from now, these disks will
have long evaporated or formed planets. New disks or cores will have obviously
formed by then but they will be at the outskirts of the 10 pc wide HII region
created by $\theta ^1$ C Ori \citep{Chevalier1999}. At such a distance, the
quantity of $^{26}$Al delivered into these disks or cores is orders of
magnitude lower than the quantity present in the Solar System
\citep{Looney2006}. In other words, supernova remnants nearby dense phases are
extremely rare \citep{Tang2014}. In conclusion, it seems very unlikely that a
nearby supernova was close enough to the Solar System to provide the known
inventory of $^{26}$Al.

If the low value of $^{60}$Fe inferred from chondrites is correct, then the
presence of $^{26}$Al remains to be explained. Supernovae can be excluded as
they would vastly overproduce $^{60}$Fe (see above). The winds of massive stars
have long been known to be $^{26}$Al and $^{60}$Fe-poor \citep{Arnould1997}. In
the model of \citet{Gaidos2009} and \citet{Young2014}, $^{26}$Al is injected at
the molecular cloud phase by the winds from a large number of  Wolf Rayet
stars. The problem with these models is that the Wolf Rayet phase is followed
by the supernova explosion and therefore they produce a large excess of
$^{60}$Fe relative to $^{26}$Al and their respective Solar System values. To
escape that caveat, \citet{Young2014} has argued that Wolf Rayet stars do not
explode as supernovae but directly collapse into black holes. Though this
possibility has been theoretically envisioned, the recent observation of a Wolf
Rayet star going supernova shows this is far from being the rule
\citep{Galyam2014}. In addition, models considering injection at the global
molecular cloud phase cannot account for the observed heterogeneity of
$^{26}$Al (see Section \ref{s:SLR}). 

The last class of models considered injection at the scale of single massive
stars. \citet{Tatischeff2010} have envisioned a single star that escaped from
its parent cluster and interacted with a neighboring molecular cloud, injecting
$^{26}$Al through its dense wind. It is compromised again by the proximity of
the Wolf Rayet phase with the supernova phase. In addition, Wolf Rayet stars
are rare. In contrast \citet{Gounelle2012} proposed that $^{26}$Al has been
injected in a dense shell of mass $\sim$1000 M$_{\sun}$ collected by the wind
of a massive star. Evolutionary models of rotating stars are used, so the
injection in the shell starts as early as the entry of the star onto the main
sequence, lasts for some Myr and ends well before the supernova explosion. When
the collected shell has become dense enough and gravitationally unstable, it
collapses and a second generation of stars form which contain $^{26}$Al.
Detailed calculations have shown that as long as the parent star, baptized {\it
Coatlicue}, is more massive than $M_{\rm min}$ = 32 M$_{\sun}$, the abundance
of $^{26}$Al in the shell is equal to or larger than that of the Solar System,
depending on the mixing efficiency of the wind material with the shell. Because
the mixing time-scale of the dense shell is comparable to its collapse
time-scale, a certain level of $^{26}$Al heterogeneity is expected
\citep{Gounelle2012} in agreement with observations. This model is in line
with observations of induced star-formation within dense shells around massive
stars \citep{Deharveng2010}. Because it corresponds to a common -- though not
universal --  mode of star formation, it implies that the Solar System is not
the only of its kind to have formed with $^{26}$Al, and that early
differentiation of planetesimals might have been common in exo-planetary
systems. \citet{Gounelle2014} has estimated that the occurrence of planetary
systems which are rich in $^{26}$Al and poor in $^{60}$Fe is on the order of
1\%. 

\subsection{Timing of planetesimal accretion}

\label{Chronology}

Planetesimal accretion itself cannot be dated \textit{directly} with
radioisotopic systems, since the mere agglomeration of different solids incurs
no isotopic rehomogenization between different mineral phases. Thus one can
only obtain upper limits with the age-dating of pre-accretionary components
(for chondrites) and lower limits from that of secondary (parent body)
processes.

Important and ever-improving constraints have emerged since the publication of
\textit{Asteroids III}. In particular, Hf-W systematics of achondrites and
irons have evidenced early differentiation, sometimes contemporaneous (within
errors) with refractory inclusion formation \citep{Kruijeretal2012}. This
indicates that planetesimal formation started very early in the evolution of
the solar system. Intriguingly, \citet{LibourelKrot2007} ascribed some olivine
aggregates in chondrites to this first generation of planetesimals \citep[but
see][ for the alternative view that these formed in the protoplanetary
disk]{Whattametal2008,Jacquetetal2012CC}. The above evidence for early
differentiation is consistent with thermal modeling expectations, as the
initial abundance of $^{26}$Al was sufficient to melt planetesimals, so that
chondrites had to be accreted later to be preserved to the present day. 

Lower limits on chondrite accretion ages may be obtained from phases
precipitated during aqueous alteration (see chapter by {\it Krot et al.}).
\citet{Fujiyaetal2013} obtained ages of 4562.5-4563.8 Myr in CI and CM
chondrites \citep[recall the accepted age of CAI formation of 4567.3$\pm$0.16
Myr, ][]{Connelly+etal2012}. As to non-carbonaceous chondrites, recent Mn-Cr
dating of fayalite formed during incipient fluid-assisted metamorphism in the
least metamorphosed LL ordinary chondrites yields an age of
4564.9$^{+1.3}_{-1.8}$ Myr \citep{Doyleetal2014}. Al-Mg systematics in the
mildly metamorphosed H chondrite Sainte Marguerite indicate an age of
4563.1$\pm$0.2 Myr \citep{ZinnerGoepel2002}. Thermal modeling based on
$^{26}$Al heating of the H chondrite parent body (one of the ordinary chondrite
groups) constrained by dating of chondrites of different metamorphic degrees
indicate accretion ages ranging from 1.8 to 2.7 Myr after refractory inclusions
\citep[see][and references therein]{Gailetal2014}, while \citet{Fujiyaetal2013}
advocate accretion of CI and CM chondrites 3-4 Myr after ``time zero'' based on
similar calculations. If $^{26}$Al really is the heat source behind chondrite
alteration, then it would indeed make sense that carbonaceous chondrites,
little affected by thermal metamorphism, accreted \textit{later} than their
non-carbonaceous counterparts \citep{GrimmMcSween1993}.

Upper limits are provided by the ages of chondrules. In both carbonaceous and
non-carbonaceous chondrites, chondrule Pb-Pb ages in individual meteorites span
a range of $\sim$4564-4567 Myr \citep{Connelly+etal2012}, i.e.\ 0-3 Myr after
refractory inclusions, with younger ages reported by Al-Mg dating for CR
\citep{KitaUshikubo2012} and enstatite chondrite \citep{Guanetal2006}
chondrules. Little correlation is seen between the age of chondrules and their
composition \citep{Villeneuveetal2012}. The significance of Al-Mg ages for
chondrules, in particular in relationship with apparently somewhat older Pb-Pb
ages \citep[e.g.][]{Connelly+etal2012}, nevertheless remains uncertain.

To summarize, it seems clear that differentiated planetesimals accreted from
the outset of the disk evolution while the known chondrites accreted later in
the evolution of the disk ($\sim$2-4 Myr after refractory inclusions), as
presumably required to escape differentiation, but the exact chronology of
chondrite formation and alteration is still in the process of being firmly
established.

\subsection{Accretion of chondrules directly after formation?}

\citet{Alexanderetal2008} proposed that the retention of sodium, a volatile
element, in chondrules during their formation indicated high solid densities in
the chondrule-forming regions, up to 7 orders of magnitude above expectations
for the solar protoplanetary disk and possibly gravitationally bound \citep[see
also][]{CuzziAlexander2006}, although quite in excess of what the proportion of
compound chondrules suggest \citep{Cieslaetal2004}. This raises the possibility
that the formation of chondrules and chondrites, respectively, may have been
contemporaneous, as also advocated by \citet{Metzler2012} based on the
existence ``cluster chondrites'' comprised of mutually indented chondrules
\citep[but see][ for a criticism of such hot accretion
models]{RubinBrearley1996}.

A further argument put forward by \cite{AlexanderEbel2012} is that chondrule
populations in different chondrite groups are quite distinct
\citep{Jones2012}. Indeed \cite{Cuzzi+etal2010} noted that two populations of
particles formed simultaneously at 2 and 4 AU would be well-mixed within 1 Myr
for $\alpha = 10^{-4}$. Here $\alpha$ is the non-dimensional measure of the
turbulent viscosity and diffusion. Relevant values for protoplanetary disks are
discussed further in Section \ref{s:streaming}, but we note here that a value
of $10^{-4}$ corresponds to the conditions which are expected if the mid-plane
in the asteroid formation region is stirred by turbulent surface layers
\citep{Oishi+etal2007}. This would suggest that chondrules had to accrete
rapidly to avoid homogenization. However, we do not know exactly the
turbulence level or original space-time separations between
chondrule-/chondrite-forming locations, as the asteroid belt may have undergone
significant reshuffling \citep{Walshetal2011}. Chondrite groups vary
significantly in bulk composition. This indicates that there has been no
thorough mixing of chondrite components, whatever their individual
transformations in the intervening time were, over the whole chondrite
formation timescale. So whatever fraction of that time the period between
chondrule formation and chondrite accretion actually represents, it is not to
be expected that chondrules should have been well-mixed over the different
chondrite-forming reservoirs \citep{Jacquet2014review}.
\begin{figure*}
  \begin{center}
    \includegraphics[width=0.45\linewidth]{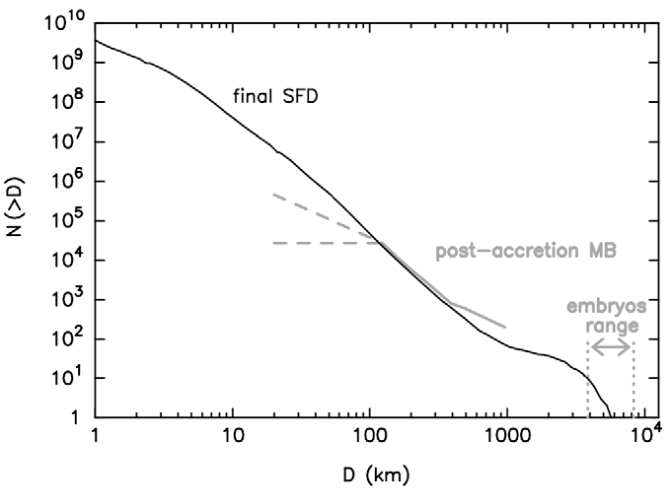}
    \quad
    \includegraphics[width=0.45\linewidth]{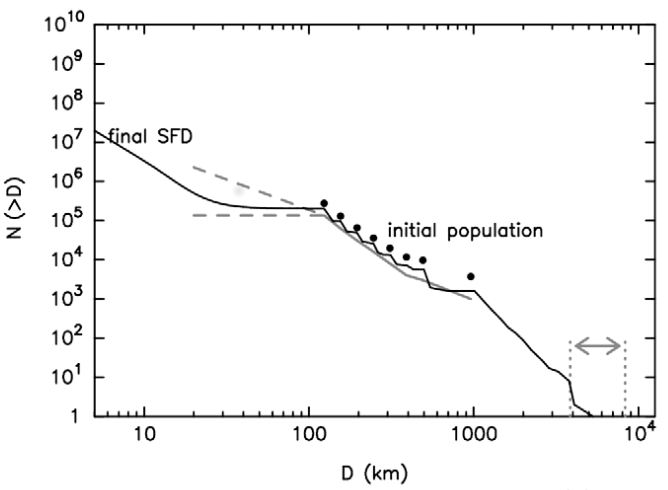}
  \end{center}
  \caption{The cumulative size distribution of asteroids $N ($$>$$D)$, as a
  function of asteroid diameter $D$, from \cite{Morbidelli+etal2009}. These
  coagulation models started with either km-sized planetesimals (left plot) or
  an initial size distribution following the current, observed size
  distribution of asteroids between 100 and 1000 km in diameter (right plot).
  The grey line shows the current size distribution of asteroids larger than
  100 km in diameter. The model with small planetesimals overproduces asteroids
  smaller than 100 km in diameter (the upper dashed line represents the current
  size distribution of small asteroids while the lower dashed lines indicates a
  tighter constraint on the size distribution directly after accretion of the
  main belt). Starting with large asteroids gives a natural bump in the size
  distribution at 100 km in diameter, as the smaller asteroids are created in
  impacts between the larger primordial counterparts.}
  \label{f:morby}
\end{figure*}

A link between chondrules and matrix is suggested, in the case of carbonaceous
chondrites, by \textit{complementarity}: the \textit{bulk} meteorite is solar
in some respect (e.g., the Mg/Si ratio) but its separate \textit{components}
(chondrules / matrix / refractory inclusions) are not \citep[chondrules have a
typically higher Mg/Si than solar while the converse is true for the matrix,
][]{HezelPalme2010}. Complementarity -- if verified, as for at least some
elements it may reflect analytical biases or parent body processes
\citep{Zandaetal2012} -- would indicate a genetic relationship between
chondrules and matrix, which would have exchanged chemical elements upon
formation, a relationship which would not be predicted e.g.\ in an X-wind
scenario for chondrule formation \citep{HezelPalme2010} in which chondrules and
matrix would have formed in different locations. But it does \textit{not}
require immediate accretion of chondrule and matrix. It only requires
chondrules and dust grains to have been transported in a statistically similar
way, as was likely the case for a large portion of the disk evolution until
accretion. Several batches of chondrule+dust may have contributed to a given
chondrite-forming reservoir provided again they suffered no loss of chondrules
relative to dust or vice-versa \citep{Cuzzi+etal2005,Jacquet+etal2012}. This
nonetheless does assume that at the stage of chondrule / matrix agglomeration,
there was no bias for or against the incorporation of any component
\citep{Jacquet2014size}, which may be an important constraint on the accretion
process. The problem is that small dust grains are much harder to incorporate
into asteroidal bodies than the macroscopic chondrules, due to their strong
frictional coupling with the gas. One could nevertheless envision that
chondrules and matrix agglomerated together as compound objects (see Section
4.2.5) prior to incorporation in asteroidal bodies and/or that matrix-sized
dust coagulated with ice into lumps with aerodynamical properties equivalent to
chondrules.

So the jury is still out on whether chondrule formation immediately preceded
incorporation in a chondrite or not. Given the chondrule age spread of 3 Myr
within individual chondrites \citep{Connelly+etal2012}, as well as the presence
of refractory inclusions and presolar grains which would not have survived
chondrule-forming events, it is possible that chondrite components did spend up
to a few Myr as free-floating particles in the gaseous disk prior to accretion.

\section{\textbf{CONSTRAINTS FROM THE ASTEROID BELT}}
\label{s:asteroids}

The modern asteroid belt contains only a fraction of its original planetesimal
population. However, the shape of the size distribution of the largest
asteroids is primordial and gives important insights into the birth sizes of
the planetesimals. Asteroid families provide a way to probe whether the
asteroids are internally homogeneous or heterogeneous on large scales.

\subsection{\bf Asteroid size distribution}

The observed size distribution of asteroids in the main belt shows a quite
steep slope for bodies with diameter $D$$>$100 km and a much shallower slope
for smaller bodies \citep{Bottke+etal2005}. A similar change of slope with an
elbow at $D$$\sim$130 km is observed in the Kuiper belt population
\citep{Fraser+etal2014}.

It was expected that the transition from a steep to a shallower slope is the
consequence of the collisional disruption of smaller bodies. However,
\cite{Bottke+etal2005} reached the opposite conclusion by examining the
collisional evolution of the asteroid belt in detail. They used a number of
constraints (the total number of catastrophic asteroid families, the survival
of the basaltic crust on Vesta, the existence of only 1-2 major basins on that
body, etc.) to conclude that the integrated collisional activity of the
asteroid belt had to be less than the one of the current main belt population
in a putative time-span of 10 Gyr. If one supposes that initially the asteroid
belt size distribution had a unique slope (the slope now observed for $D$$>$100
km), such a limited collisional evolution is not sufficient to reduce the slope
of the size distribution of objects smaller than 100 km down to the observed
value, i.e.\ to create the observed elbow. Therefore, \cite{Bottke+etal2005}
concluded that the elbow at $D$$\sim$100 km is a fossil feature of the
primordial size distribution. For the Kuiper belt, the constraints on the
integrated collisional activity are not as tight as for the asteroid belt.
Nevertheless, models seem to suggest that collisional evolution alone could not
create an elbow at diameters larger than 80 km \citep{PanSari2005}, which is
significantly smaller than the observed value.

\cite{Morbidelli+etal2009} failed to produce the elbow at $D$$\sim$100 km in
the asteroid belt in collisional coagulation simulations starting from a
population of small planetesimals (see Figure \ref{f:morby}). So, having in
mind the new models of formation of large planetesimals from self-gravitating
clumps of chondrules or larger pebbles and boulders
\citep{Johansen+etal2007,Cuzzi+etal2008}, they proposed that 100 km was the
minimal diameter of the original planetesimals. Moreover, not being able to
reproduce the current slope of $D$$>$100 km asteroids by mutual collisions
between bodies of 100 km in size, \cite{Morbidelli+etal2009} argued that these
large planetesimals were born with a similar slope. However, as we will see in
Section \ref{s:layered}, the current slope can be reproduced by considering the
accretion of chondrule-sized particles by 100-km-scale planetesimals during the
gaseous disk phase, a process not considered by \cite{Morbidelli+etal2009}.

\cite{Weidenschilling2011} managed to reproduce the elbow at $D$$\sim$100 km in
the asteroid belt from collisional coagulation simulations starting from
objects of 50 - 200 m radius. Because of the small size of these objects,
collisional damping and gas drag keep the disk very dynamically cold (i.e.
with a small velocity dispersion among the objects). Hence, in the simulations
of \cite{Weidenschilling2011}, the elbow at $D$$\sim$100 km is produced by a
transition from dispersion-dominated runaway growth to a regime dominated by
Keplerian shear, before the formation of large planetary embryos. However, any
external dynamical stirring of the population, for instance due to gas
turbulence in the disk, would break this process. Moreover, these simulations
are based on the assumption that any collision which does not lead to
fragmentation results in a merger, but 100-m-scale objects have very weak
gravity and the actual capability of bodies so small to remain bound to each
other is questionable. Finally we stress that the formation of 100-m-scale
bodies is an open issue, in view of the bouncing barrier and meter-size barrier
discussed in Section \ref{s:dust_growth}.

\subsection{\bf Snowline problems}

Among the various meteorite types that we know, carbonaceous meteorites (or at
least some of them like CI and CM) contain today a considerable amount (5-10\%)
of water by mass. Evidence for water alteration is wide-spread, and it is
possible that the original ice content of these bodies was higher, close to the
50\% value expected from unfractionated solar abundances. Instead, ordinary
chondrites contain $<2$\% water by weight
\citep[e.g.][]{Jarosewich_1990,Krot2009}; while ordinary chondrites do show
signs of water alteration, it is unlikely that they ever contained 50\% water
ice by mass. This suggests that the parent bodies of these meteorites (CI and
CM vs.\ ordinary) formed on either side of the condensation line for water,
also called the {\it snowline}. In other words, the snowline was located in the
middle of the asteroid belt at the time when the asteroids formed. The Earth is
also very water poor (some 0.1\% by mass, although uncertain by a factor of
$\sim$5), suggesting that the planetesimals in its neighborhood were mostly dry
and that water was delivered only by the small amount of planetesimals accreted
from farther out \citep{Morbidelli+etal2000, Raymond+etal2004,
OBrien+etal2014}. The very dry enstatite chondrites have been proposed to arise
from an extinct portion of the asteroid belt located between 1.6 and 2.1 AU
from the Sun \citep{Bottke+etal2012}.

The problem with this picture is that chondritic bodies accreted late (2-4 Myr
after CAIs) and that the snowline is expected to migrate towards the Sun with
time. The temperature of the disk is set by the equilibrium between heating and
cooling. In the inner part of the disk, the heating is predominantly due to the
viscous friction of the gas in differential rotation around the Sun, so it is
related to the gas accretion rate of the star. The cooling rate is governed by
how much mass is in the disk in the form of micron-sized grains. In
\cite{Bitsch+etal2014c}, the snowline is at 2-3 AU only in the earliest phases
of disk evolution when the accretion rate is $\dot{M}=10^{-7} M_\odot\,{\rm
yr}^{-1}$; but the snowline migrates to 1-2 AU already after 500,000 yr when
the accretion rate drops below $\dot{M}=10^{-8} M_\odot\,{\rm yr}^{-1}$. These
models are probably optimistically warm, because it is assumed that the ratio
between the mass in $\mu$m-sized dust and the mass in gas is 1\% (this reflects
the solar metallicity but planetesimal accretion should eventually decrease the
dust content) and account for the heating produced by stellar irradiation by a
star more luminous than our early Sun. According to the observation of the
accretion rate of stars as a function of age \citep{Hartman+etal1998} and
photoevaporation models \citep{AlexanderArmitage2006}, the accretion rate on
the star should drop to $\dot{M}=10^{-9} M_\odot\,{\rm yr}^{-1}$ at 2-3 Myr,
i.e.\ at the time of chondrite formation. The fact that there is no sign of
primary accretion in the Solar System after $\sim$3 Myr also suggests that the
disk disappeared at that time (or that all the solids could have been
incorporated into parent bodies at that time -- however, accretion of solids is
unlikely to reach 100\% efficiency). Thus, at the time of chondrite formation,
the snowline should have been well inside the inner edge of the asteroid belt,
possibly even inside 1 AU. This is at odds with the paucity of water in the
ordinary chondrites and in the Earth.

A way to keep the disk warm has been proposed recently by
\cite{MartinLivio2012,MartinLivio2013}. In their model, a dead zone in the
disk, with low turbulent viscosity, becomes very massive and develops
gravitational instabilities which heat the gas. The unlimited pile-up of gas in
the dead zone, however, could be a consequence of a simplified disk transport
model, and is not observed in more complex 3D hydrodynamical calculations
\citep{Bitsch+etal2014b}. Also, the model of \cite{MartinLivio2012} predicts a
cold region where ice could condense inside the orbit of the Earth, which seems
inconsistent with the compositions of Venus and Mercury.

A possible way to keep the asteroid belt ice-poor, despite freezing conditions,
is to stop the inwards flow of icy pebbles in a pressure bump. The inwards
drift of pebbles is caused by friction with the slower moving gas (this effect
is discussed more in Section \ref{s:si}). Then when the nebula cooled locally
below the ice condensation temperature, there was no water vapor present to
condense, and the planetesimals formed there would have a volatile depletion
similar to that achieved in a warm disk. \cite{Bitsch+etal2014b} explored the
effects of viscosity transitions at the snowline \citep[as suggested
by][]{KretkeLin2007}, but found that only very steep radial gradients in the
$\alpha$ parameter allow the formation of a pressure bump.
\cite{Lambrechts+etal2014} showed that the formation of a proto-Jupiter of
about 20 M$_{\rm E}$ can produce a strong pressure bump just exterior to its
orbit. If proto-Jupiter formed at the snowline when the snowline was at the
outer edge of the asteroid belt, then its presence would have shielded the
asteroid belt and the terrestrial planet region from the flow of icy pebbles,
while small silicate dust and chondrules would remain for longer times in the
asteroid belt due to their slow radial drift.

\subsection{Are asteroids internally homogeneous?}
\label{s:homo}

The possibility that the asteroids grew incrementally by layered accretion of
chondrules (Section \ref{s:layered}) implies that asteroids are internally
heterogeneous, in the most extreme case with a chondritic surface layer
residing on a differentiated interior. We discuss here briefly whether such
internal heterogeneity is supported by observations of the asteroid belt.
\citet{Burbine_etal_2002} note that 100-150 distinct meteorite parent bodies,
3/4 of them differentiated, are represented in the meteorite collection.
However, this sample is biased towards the sturdy irons and against the weaker,
never-melted primitive chondrites.

Differentiation is a constraint on formation age. Most studies have suggested
that sizeable asteroids forming $<10^6$ years after CAIs are almost certain to
have thoroughly melted, but those which formed more than 2 Myr after CAIs may
have escaped melting except near their centers. More recent work paints a more
complicated picture (see chapter by {\it Elkins-Tanton et al.}). Specifically,
the Allende parent body, source of primitive CV3 chondrites, is thought to have
melted near its center, as evidenced by the paleomagnetism detected in these
meteorites \citep{Elkins-Tanton+etal2011}; other CV chondrites may show a
similar signal \citep{Weiss_etal_2010_SSR}. This suggests that the CV parent
body was differentiated in its interior, but preserved an undifferentiated
chondritic layer. Other authors have ascribed the magnetic field in the CV
chondrites to impacts \citep[e.g.][]{Wasson+etal2013}.

Modeling of the buoyancy of silicate melts of different composition suggests
that, for C-type composition, melt might be dense and remain stable at depth,
but for S-type (OC) compositions, and certainly for Enstatite compositions,
melt is less dense than surrounding material and will rise to manifest on the
surface \citep{Fu_Elkins-Tanton_2014}. Thus there might be old,
centrally-melted C-type asteroids such as the Allende parent body, where the
evidence for differentiation remains buried, but the lack of evidence for
significant surface melting on most S-type asteroids may argue that most of
them remain unmelted and undifferentiated throughout \citep[see chapter by {\it
Elkins-Tanton et al.}, and][]{Weiss_Elkins-Tanton_2013}.

We can sample the internal properties of asteroids in two ways - from the
observed color and albedo distributions of collisionally disrupted asteroids in
families (see also chapters by {\it Nesvorny et al.}\ and {\it Michel et al.})
and from the meteorites that derive from them (see next paragraph). The NEOWISE
mission has measured the albedos of a large number of asteroids and families,
finding a dichotomy in albedo, roughly corresponding to the classical S and C
types, that is most evident in the outer main belt. Collisional families cover
the entire belt, so they avoid the sampling bias that affects meteorites. Many
families have internal albedo distributions that are narrower than the global
spread of albedos {\it across} families. A similar story is told by
observations at visual and near-IR wavelengths
\citep{Mothe-Diniz_Nesvorny_2008}. This conclusion is nevertheless complicated
by the identification of interlopers in (and exclusion from) the asteroid
families. Hence there is an inherent tendency to observe low albedo variations
within asteroid families. In contrast, the Eos family and the Eunomia family
have unusually large internal variance \citep{Mothe-Diniz_etal_2005,
Mothe-Diniz_etal_2008} and the Eunomia family looks like what an internally
differentiated S-type might produce \citep[see discussion
in][]{Weiss_Elkins-Tanton_2013}.

The three ordinary chondrite groups (H, L, and LL) are each thought to derive
from a single parent body, based on a clustering of cosmic ray exposure ages of
chondrites in each group, as if they were excavated by the same few large
impacts \citep{Eugster_etal_2006}. The thermal history of the H chondrite
parent body under internal heating by $^{26}$Al was modeled by
\citet{Trieloff_etal_2003}, \citet{Monnereau_etal_2013}, and
\citet{Henke_etal_2013} who all concluded it was a roughly 100-km-radius body.
Chondrites from all three OC groups show different amounts of thermal
alteration from this process, designated as metamorphic grades 3-6. Models
imply that the H chondrite parent, at least, first incurred thermal alteration
in an ``onion-shell" fashion, with the most strongly heated H5/H6 material
heated deep in the interior, and then was catastrophically disrupted and
reassembled as a rubble pile \citep[Figure 1,][]{Taylor_etal_1987,
Scott_etal_2014}. The small fractional abundance of minimally altered
chondrites (H3/H4; Figure 1) constrains the accretion of the H parent body to
have happened quickly - probably faster than $3 \times 10^5$ yr
\citep{Henke_etal_2013, Vernazza_etal_2014}). The constant mass growth rate
adopted in \cite{Henke_etal_2013} is nevertheless not applicable to layered
accretion, which results in run-away accretion and deposition of most of the
mass towards the end of the growth phase (see further discussion in Section
\ref{s:layered}).

While the major element {\it chemical} compositions, and the {\it oxygen
isotopic} compositions, of the OC groups differ significantly, there is little
or no discernible variation of either chemical or isotopic composition across
metamorphic grades in any of the three groups \citep{Jarosewich_1990,
Wood_2005}; see also Tables 1 and 2 of \citet{Clayton_etal_1991}. The
variation of {\it chondrule size} across metamorphic grade is less well studied
(A. Rubin, personal communication 2014).

\cite{Dodd1976} demonstrated that the difference in the aerodynamical friction
time between metal grains and silicate chondrules can explain silicate-metal
fractionation in the ordinary chondrites. In this picture the LL chondrites are
under-abundant in metal because the parent body was successfully able to
accrete large silicate-rich chondrules \citep[the chondrules in the LL
chondrites are larger than those in both H and L chondrites, see
e.g.][]{NelsonRubin2002}. The oxidation of metallic Fe could have happened as
the LL chondrite parent body accreted water-bearing phyllosilicates
\citep{Rubin2005}. In the layered accretion model presented in Section
\ref{s:layered}, asteroids will accrete larger and larger particles as they
grow. Metal grains had a more restricted size range and hence a relatively
small parent body of the H chondrites would accrete mainly small chondrules
together with metal grains. A small size of the H chondrite parent body is
further supported by the low fraction of strongly metamorphosed samples that we
have from that group \citep{Dodd1976}. A similar story is told for the
enstatite chondrites \citep{Schneideretal1998}: the EH group has more metal and
smaller chondrules than the EL group. This aerodynamical size sorting may be
evidence of asteroid growth by layered accretion (Section \ref{s:layered}) or
asteroid formation by turbulent concentration (Section
\ref{s:turbulent_concentration}).

\section{\textbf{DUST GROWTH BY STICKING}}
\label{s:dust_growth}

Now that we have given an overview of some of the constraints from meteorites
and asteroids, we can turn to the theoretical models of planetesimal
formation. In this section we discuss the growth of dust by direct sticking;
subsequent chapters discuss gravitational instability models.

\subsection{\bf Particle-gas interaction}

The dynamical behavior of a particle in gas depends on both its size and
density, as determined by its {\it friction time} $\tau_{\rm f}$ in the nebula
gas \citep{Whipple1972,Weidenschilling1977}. For particles smaller than the
gas molecule mean free path (approximately 10-100 cm in the asteroid belt
region, depending on the uncertain value of the gas density), the friction time
is
\begin{equation}\label{ts}
  \tau_{\rm f} = \frac{a \rho_\bullet}{v_{\rm th} \rho_{\rm g}} \, ,
\end{equation}
where $a$ and $\rho_\bullet$ are the particle radius and density, and $v_{\rm
th}$ and $\rho_{\rm g}$ are the gas thermal speed and mid-plane density. The
thermal speed of the gas molecules is in turn connected to the isothermal sound
speed through $v_{\rm th}=\sqrt{8/\pi} c_{\rm s}$. The Stokes number ${\rm St}$
is defined as ${\rm St} = \varOmega \tau_{\rm f}$, where $\varOmega$ is the
(local) orbital frequency of the protoplanetary disk. The translation from
Stokes number to particle size follows
\begin{equation}\label{St}
  a = \frac{(2/\pi)\varSigma_{\rm g} {\rm St}}{\rho_\bullet} \approx
  80\,{\rm cm}\,{\rm St}\,f_{\rm g}(r) \left( \frac{r}{2.5\,{\rm AU}}
  \right)^{-1.5} \, .
\end{equation}
Here $\varSigma_{\rm g}=\sqrt{2 \pi} \rho_{\rm g} H_{\rm g}$ is the gas column
density and $f_{\rm g}(r)$ a parameter which sets gas depletion relative to the
MMSN as a function of semi-major axis $r$. We set in the second equality
$\rho_\bullet=3.5\,{\rm g/cm^3}$, relevant for chondrules. The Stokes number
controls many aspects of dust dynamics. Particles of larger Stokes numbers
couple increasingly to larger, longer-lived, and higher-velocity eddies in
nebula turbulence, thus acquiring larger relative velocities. Solutions for
particle velocities have been developed by \citet{Voelk_etal_1980} and
\citet{Mizuno_etal_1988}, including closed-form analytical expressions by
\citet{Cuzzi_Hogan_2003} and \citet{Ormel_Cuzzi_2007}. Importantly, the Stokes
number also controls the degree of sedimentation, with the scale-height of the
sedimented mid-plane layer, $H_{\rm p}$, given by
\begin{equation}
  H_{\rm p} = H_{\rm g} \sqrt{\frac{\alpha}{{\rm St}+\alpha}} \, .
\end{equation}
Here $H_{\rm g}$ is the gas scale-height and $\alpha$ is a measure of the
turbulent diffusion coefficient $D$ normalised as $D=\alpha c_{\rm s} H_{\rm
g}$ \citep[see][ for references]{Johansen+etal2014}. Significant sedimentation
can occur when ${\rm St} \gtrsim \alpha$. Chondrules of mm sizes have a Stokes
number of ${\rm St}\sim10^{-3}$; hence chondrules will only settle out of the
gas if $\alpha \ll 10^{-3}$.

\subsection{\bf Dust growth}

The study of dust growth has been an extremely active field, both
experimentally and numerically, since {\it Asteroids III}. Subsequent reviews
were presented by \citet{Dominik_etal_2007}, \citet{Blum_Wurm_2008} and
\cite{ChiangYoudin2010}; the older review by \citet{Beckwith_etal_2000} is also
valuable for basics. Two very recent overview chapters presenting both the
basic physics and selected recent highlights are by \citet{Johansen+etal2014}
and \citet{Testi_etal_2014}; for efficiency we build on those chapters and here
emphasize specifics relevant to asteroid formation. 

\subsubsection{Sticking}

Particles can stick if their relative kinetic energy exceeds certain functions
of the surface energy of the material, which depends on composition. At the low
relative velocities for small monomers (0.1-10 $\mu$m) under nebula conditions,
both ice and silicate particles stick easily and form loose, porous aggregates.
The process continues until the aggregates are at least 100 $\mu$m in radius.
Aggregates can continue to grow and stick at larger velocities, if their open
structure is able to deform and dissipate energy \citep{Wada_etal_2009}. The
entire process of growth beyond roughly 100 $\mu$m fluffy aggregates depends on
just how much these aggregates can be compacted by their mutual collisions.
Recent studies that concentrate on icy particles outside the snowline have
argued that the high surface energy of ice prevents significant compaction from
occurring (and keeps relative velocities small) until particles have grown to
extremely large size - hundreds of meters - with extremely low density
\citep{Okuzumi_etal_2012}.

\subsubsection{Bouncing}

The surface energies of silicates are much smaller than those of ice, so it is
easy for even mm-sized silicate particles to compact each other in mutual
collisions. Relative velocities large enough to cause compaction and bouncing
are acquired by roughly mm-size silicate particles in nebula turbulence.
Coagulation modeling by \citet{Zsom_etal_2010}, consistent with experiments
\citep{Guettler+etal2010,Weidling_etal_2012}, revealed a {\it bouncing barrier}
in this size range where growth of silicate aggregates by sticking ceased. This
new barrier joins the long-known fragmentation barrier and radial drift barrier
which, even if the bouncing barrier can be breached, tend to frustrate growth
in the asteroid belt region beyond dm-m size \citep{Brauer+etal2008a,
Birnstiel+etal2010, Birnstiel_etal_2011}.

\subsubsection{Fragmentation}

A simple critical velocity $v_{\rm frag}$ can be used to refer to fragmentation
of two comparable masses. This approach has been modified in some treatments to
include some {\it mass transfer} by a smaller projectile hitting a larger
target at high velocity, even if the projectile is destroyed and some mass is
ejected from the target \citep{Wurm+etal2005}. An alternate approach is to
treat the fragmentation threshold as a critical kinetic energy per unit mass
$Q^*$, which can be thought of as a critical velocity squared for same-size
particles \citep{Stewart_Leinhardt_2009}. The latter treatment automatically
accounts for particle size differences and thus allows accretion of small
particles to proceed at collision velocities much higher than the nominal
$v_{\rm frag} \sim \sqrt{Q^*}$. \citet{Stewart_Leinhardt_2009} treated solids
as weak rubble piles, all calibrated using experimental work by
\citet{Setoh_etal_2007}. These expressions allow for the higher efficiency of
low-velocity collisions in fragmentation than for hypervelocity impacts. For
particles made of small silicate grains, a value of $Q^*$ on the order of
$10^4$ cm$^2$/s$^2$ is suggested, with an associated $v_{\rm frag} \sim $1 to
several m/s, for weak cm-m-sized aggregates \citep{Schraepler_etal_2012,
Stewart_Leinhardt_2009, Wada_etal_2009, Wada_etal_2013} or even less
\citep{Beitz_etal_2011}.

\subsubsection{Lucky particles}

The bouncing barrier, in preserving most of the available solids at small
sizes, may provide a target-rich environment for growth of much larger ``lucky"
particles, which experience few, or low-velocity, collisions and avoid
destruction while steadily growing from much smaller particles to large sizes
\citep{Windmark+etal2012a,Windmark+etal2012b,Garaud+etal2013,Drazkowska+etal2013}.
However, \cite{Windmark+etal2012a,Windmark+etal2012b} and
\cite{Garaud+etal2013} did not include radial drift, which is important because
the growth time for these lucky particles greatly exceeds the drift time to the
Sun. \cite{Drazkowska+etal2013} removed the radial drift problem with a
pressure bump at the edge of a dead zone, and still found that the total number
of meter-size particles they could produce in $3 \times 10^4$ years was in the
single digits. An isolated particle cannot trigger collective effects such as
the streaming instability (Section 6), but can only keep growing by sweep up.
\cite{Johansen+etal2008} and \cite{Xie_etal_2010} have modeled this ``snowball"
stage and find that growth in this fashion is extremely slow unless the nebula
turbulent $\alpha$ is very low, because the small feedstock particles are
vertically diffused to a low spatial density otherwise. 

\subsubsection{Chondrule rims and chondrule aggregates}

The fine-grained component of chondrites is not only found in a featureless
background matrix; it is also found rimming individual chondrules and other
coarse particles, often filling cavities \citep{Metzler_etal_1992,
Brearley1993, Cuzzi+etal2005}. The origin of these {\it fine grained rims} has
been debated \citep{Lauretta_etal_2006}. One school of thought regards them as
accretionary rims, swept up as a cooled chondrule moves relative to the gas and
entrained dust or small aggregates \citep{MacPherson_etal_1985,
Metzler_etal_1992, Bland_etal_2011}. Matrix and rims were reviewed in depth by
\citet{Scott_etal_1984}; amongst many interesting results, they found evidence
that rims were accreted as numerous aggregates of variable mean composition,
rather than as monomers. \cite{Rubin2011} suggested that carbonaceous
chondrites formed in dusty regions of the solar protoplanetary disk and that
matrix accumulated into mm-cm-sized highly porous dust balls. In this picture,
chondrules acquired matrix rims by collisions with these dust balls rather than
in collisions with smaller particles.

The accretion of fine grained rims was modeled by \citet{Morfill_etal_1998},
\citet{Cuzzi_2004}, and \citet{Ormel+etal2008}. In these models, relative
velocities predicted for the nebula environment have been shown to be
compatible with sticking and compaction under the theory of
\citet{Dominik_Tielens_1997}; \citet{Ormel+etal2008} added the effects of
interchondrule collisions on compacting the rims further.
\citet{Ormel+etal2008} found that chondrules which had accreted porous rims of
dust between collisions could stick more easily because the porous rims acted
as shock absorbers, resulting in composite, dm-size objects formed of rimmed
chondrules.

\subsubsection{Gravitational scattering barrier}

There is one further barrier for particles trying to grow incrementally, by
sticking, to km and larger sizes. This {\it gravitational scattering} barrier
arises because, in turbulence, nebula gas has small density fluctuations
associated with pressure and vorticity. These density fluctuations, primarily
on large scales, scatter growing planetesimals to achieve high relative speeds
\citep{Laughlin+etal2004, NelsonPapaloizou2004, Nelson_Gressel_2010,
Yang+etal2012}. The random velocities acquired by 1-10-km-scale planetesimals
in this way are sufficient to put them into an erosive, rather than
accretionary, regime throughout most of the Solar System for a range of the
most plausible $\alpha$ values \citep{Ida_etal_2008, Stewart_Leinhardt_2009}.
\citet{Ormel_Okuzumi_2013} found that planetesimals need to have radii of 100
km or more to be able to survive these destructive encounters. 

Overall, the barriers to formation of 100-km-scale asteroids by incremental
growth-by-sticking appear formidable. For these reasons, a number of models
have arisen which avoid these problems with a {\it leapfrog} process in which
10-1000-km-scale asteroids form directly from smaller particles stuck at one of
these barriers.

\section{\textbf{TURBULENT CONCENTRATION}}
\label{s:turbulent_concentration}

The leapfrog model sometimes referred to as {\it turbulent concentration} (TC)
or {\it turbulent clustering} can in principle make 10-100-km-radius
planetesimals directly from chondrule-sized particles. The turbulent
concentration model was motivated originally by laboratory and numerical
experiments that showed dense particle clusters forming spontaneously in
isotropic turbulence, in which the most intense clustering was seen for
particles with friction time ${\tau_{\rm f}}$ equal to the Kolmogorov (or
smallest eddy) timescale
\citep{SquiresEaton1990,SquiresEaton1991,Wang_Maxey_1993, Eaton_Fessler_1994}.
It was realized that this condition was very closely satisfied under canonical
nebula conditions by the very chondrule-size particles that make up the bulk of
primitive chondrites \citep{Cuzzi+etal1996, Cuzzi+etal2001}.

\subsection{\bf A brief history of TC models}

What remains uncertain and a subject of current study, as with many properties
of turbulence, is just how the laboratory experiments and numerical simulations
translate into actual nebula conditions. The flow regime is described by its
turbulence Reynolds number ${\rm Re} = UL/\nu = \alpha c_{\rm s} H_{\rm g}/\nu$
where $U$ and $L$ are some typical velocity and length scale of the flow, and
$\nu$ is the molecular viscosity. All current experiments and simulations are
run at ${\rm Re}$ far smaller than likely nebula values.
\citet{Cuzzi+etal2001} noticed that the volume fraction of dense clumps
increased with increasing ${\rm Re}$, and suggested a scaling which would map
the behavior to the much higher values of ${\rm Re}$ relevant for turbulence in
protoplanetary disks.

\citet{Hogan_Cuzzi_2007} showed that mass-loading feedback of the particle
burden on gas turbulence caused the concentration process to saturate at a mass
loading $\Phi = \rho_p/\rho_g \sim 100$, precluding the small, dense clumps
advocated by \citet{Cuzzi+etal2001}. Based on this, \citet{Cuzzi+etal2008}
advocated gravitational binding and ultimate sedimentation of much larger
clumps, on the order of $10^3-10^4$ km, with $\Phi \sim 10-100$. They showed
that a long-neglected discovery by \citet{Sekiya_1983} precludes genuine,
dynamical-timescale collapse for dense clumps of chondrule-size particles under
plausible conditions. Particles of these sizes and friction times can only
sediment slowly towards the center of their bound clump, on a much longer
timescale ($10^2-10^3$ orbits). \citet{Cuzzi+etal2008} suggested a criterion
for stability of these bound, yet only slowly-shrinking, clumps which led
directly to the conclusion that they would preferentially form {\it large
planetesimals} comparable in size with the mass-dominant 100 km radius mode
advocated for the early asteroid belt by \citet{Bottke+etal2005} -- thus
leaping over the long-troublesome mm-m-km-size barriers. Objects formed in this
way are compatible with the “primary texture” seen in the most primitive
unbrecciated CM and CO chondrites \citep{Metzler_etal_1992,Brearley1993}, and
would display deep homogeneity of bulk chemical and isotopic properties.

\citet{Cuzzi+etal2010} and \citet{Chambers_2010} went on to describe an
end-to-end primary accretion scenario, combining stability thresholds with
calculated probability distributions of clump density, finding that a range of
nebula conditions (all implying $>10\times$ local enhancement of the usually
assumed 1\%  cosmic solids-to-gas ratio within some few $10^4$ km of the
midplane) could match the required rate of planetesimal formation and the
characteristic mass mode around 100-200 km diameter. \citet{Cuzzi+etal2010}
gave a number of caveats regarding the built-in assumptions of this model; one
caveat regarding scale-dependence of the concentration process has been found
to be important enough to change the predictions of the scenario quantitatively
(see below). Subsequently, \citet{Cuzzi_Hogan_2012} resolved a discrepancy in a
key timescale between \citet{Cuzzi+etal2010} and \citet{Chambers_2010}, which
makes planetesimal formation 1000 times faster than in \citet{Cuzzi+etal2010}
\citep[and correspondingly slower than in][]{Chambers_2010}. 

\subsection{\bf New insights into turbulent concentration}

The primary issues are whether  it is {\it always} Kolmogorov friction time
particles that are most effectively concentrated, {\it and} whether the physics
of their concentration is scale-invariant. \citet{Hogan_Cuzzi_2007} argued by
analogy with the observed scale-invariance of turbulent dissipation, which is
dominated by Kolmogorov-time vortex tubes (little tornados in turbulence) that
the concentration of Kolmogorov-friction-time particles would also be
scale-invariant (see also \citet{Cuzzi_Hogan_2012}. They developed a so-called
{\it cascade model} by which to extend the low-${\rm Re}$ results to nebula
conditions.

The primary accretion scenarios of \citet{Cuzzi+etal2010}  and
\citet{Chambers_2010} used this cascade model to generate density-vorticity
PDFs as a function of nebula scale. \citet{Pan+etal2011} ran simulations at
higher ${\rm Re}$ than \citet{Hogan_Cuzzi_2007} and found that the clump
density PDFs dropped faster than would be predicted by the scale-invariant
cascade. They suggested that the physics of particle concentration might
indeed {\it be} scale-dependent, and that planetesimal formation rates obtained
using the \citet{Hogan_Cuzzi_2007} cascade might be significantly
overestimated. 

Ongoing work supports this concern about scale dependence.
\citet{Cuzzi_etal_2014} have analyzed much higher ${\rm Re}$ simulations
\citep{Bec_etal_2010} and found that the cascade measures called ``multiplier
distributions", which determine how strongly particles get clustered at each
spatial scale, do depend on scale at least over the largest decade or so of
length scale; that is, the scale-invariant inertial range for particle
concentration and dissipation does not become established at the largest scale,
causing little concentration to occur until roughly an order of magnitude
smaller scale. Because the cascade process is multiplicative, this slow start
means that fewer dense zones are to be found at any given scale size than
previously thought. 

New, scale-dependent cascades can now be implemented using the results of
\citet{Cuzzi_etal_2014} at the largest scales. The quantitative implications
are not clear as of this writing. One qualitative result is that particles with
friction times 2-10$\times$ longer than originally suggested are more strongly
clustered at the most relevant (larger) length scales \citep[consistent with
previous work by][]{Bec_etal_2007,Pan+etal2011}. Because we know the size and
density of chondrules, the implications will be different preferred values of
gas density and/or turbulent $\alpha$ \citep[see][]{Cuzzi+etal2001}. A related
implication is that concentrated particle sizes may not be as narrowly peaked
as shown in \citet{Cuzzi+etal2001}. However, at the same time, new observations
are also showing broader chondrule size distributions in chondrites
\citep{Fisher_etal_2014}.

\section{\textbf{PRESSURE BUMPS AND STREAMING INSTABILITY}}
\label{s:streaming}

The turbulent concentration mechanism described in the previous section
operates on the smallest scales of the turbulent flow (although the vortical
structures which expel particles can be very elongated). The dynamical
time-scales on such small length-scales are much shorter than the local orbital
time-scale of the protoplanetary disk. In contrast, the largest scales of the
turbulent flow are dominated by the Coriolis force, and this allows for the
emergence of large-scale {\it geostrophic structures} (high-pressure regions in
perfect balance between the outwards-directed pressure gradient force and the
inwards-directed Coriolis force).

\cite{Whipple1972} found that particles are trapped by the zonal flow
surrounding large-scale pressure bumps. Pressure bumps \citep[in a way
azimuthally extended analogs to the vortices envisioned
in][]{BargeSommeria1995} can arise through an inverse cascade of magnetic
energy \citep{Johansen+etal2009a,Simon+etal2012,Dittrich+etal2013} in
turbulence driven by the magnetorotational instability
\citep{BalbusHawley1991}. Pressure bumps concentrate primarily large (0.1-10 m)
particles which couple to the gas on an orbital time-scale
\citep{Johansen+etal2006}, reaching densities at least 100 times the gas
density which leads to the formation of 1000-km-scale planetesimals
\citep{Johansen+etal2007,Johansen+etal2011}. The magnetorotational instability
is nevertheless no longer favoured as the main driver of angular momentum
transport in the asteroid formation region of the solar protoplanetary disk,
since the ionisation degree is believed to be too low for coupling the gas to
the magnetic field \citep[see review by][]{Turneretal2014}.

The magnetorotational instability can still drive turbulence (with $\alpha$ in
the interval from $10^{-3}$ to $10^{-2}$) in the mid-plane close to the star
(within approximately 1 AU where the ionisation is thermal) and far away from
the star (beyond 20 AU where ionising cosmic rays and X-rays penetrate to the
mid-plane). Accretion through the ``dead zone'', situated between these regions
of active turbulence, can occur in ionised surface layers far above the
mid-plane \citep{Oishi+etal2007}, from disk winds \citep{BaiStone2013} and by
purely hydrodynamical instabilities in the vertical shear of the gas
\citep{Nelson+etal2013} or radial convection arising from the subcritical
baroclinic instability \citep{KlahrBodenheimer2003,LesurPapaloizou2010}. The
mid-plane is believed to be stirred to a mild degree by these hydrodynamical
instabilities or by perturbations from the active layers several scale-heights
above the mid-plane, driving effective turbulent diffusivities in the interval
from $10^{-5}$ to $10^{-3}$ in the mid-plane. The inner and outer edges of this
``dead zone'', where the turbulent viscosity transitions abruptly, are also
possible sites of pressure bumps and large-scale Rossby vortices which feed off
the pressure bumps \citep{Lyra+etal2008b,Lyra+etal2009b}.

\subsection{Streaming instability}
\label{s:si}

The low degree of turbulent stirring in the asteroid formation region also
facilitates the action of the streaming instability, a mechanism where
particles take an active role in the concentration process
\citep{YoudinGoodman2005,YoudinJohansen2007,JohansenYoudin2007}. The
instability arises from the speed difference between gas and solid particles.
The gas is slightly pressure-supported in the direction pointing away from the
star, due to the higher temperature and density close to the star, which mimics
a reduced gravity on the gas. The result is that the gas orbital speed is
approximately 50 m/s slower than the Keplerian speed at any given distance from
the star. Solid particles are not affected by the global pressure gradient --
they would move at the Keplerian speed in absence of drag forces, but drift
radially due to the friction from the slower moving gas. The friction exerted
from the particles back onto the gas leads to an instability whereby a small
overdensity of particles accelerates the gas and diminishes the difference from
the Keplerian speed. The speed increase in turn reduces the local headwind on
the dust. This slows down the radial drift of particles locally, which leads to
a run-away process where isolated particles drift into the convergence zone and
the density increases exponentially with time. This picture is a bit simplified
as \cite{YoudinGoodman2005} and \cite{Jacquet+etal2011} showed that the
streaming instability operates only in presence of rotation, i.e.\ the
instability relies on the presence of Coriolis forces. This explains why the
instability occurs on relatively large scales of the protoplanetary disk where
Coriolis forces are important, typically a fraction of an astronomical unit,
and operates most efficiently on large particles with frictional coupling times
around 1/10 of the orbital time-scale (typically dm sizes at the location of
the asteroid belt).
\begin{figure*}[!t]
  \begin{center}
    \includegraphics[width=0.7\linewidth]{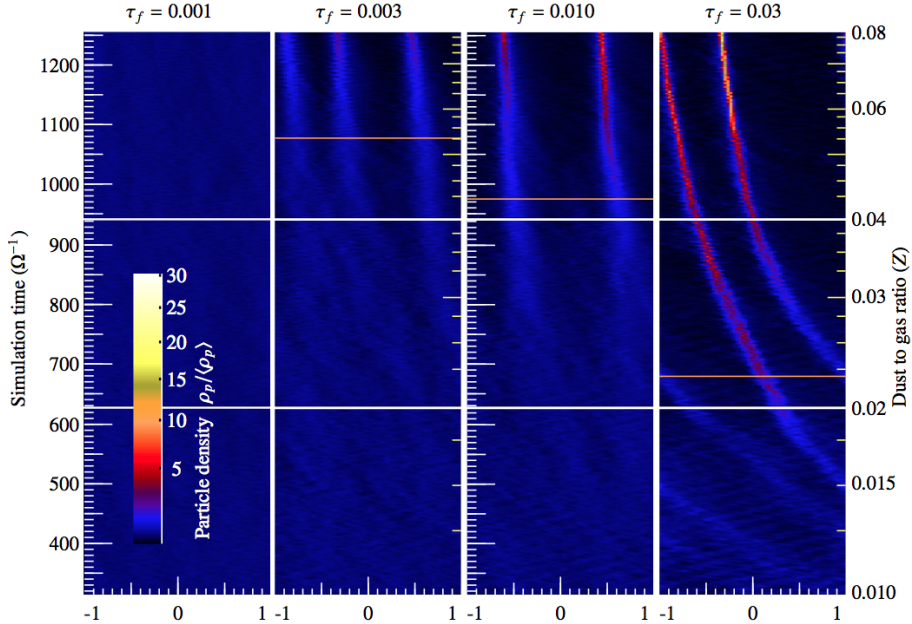}
  \end{center}
  \caption{Space-time plots of particle concentration by streaming
  instabilities, from \cite{Carrera+etal2015}, with the $x$-axis indicating the
  radial distance from the centre of the simulation box and the $y$-axis the
  time (on the left) and the dust-to-gas ratio (on the right). The four columns
  show particle sizes 0.5 mm ($\tau_{\rm f}=0.001 \varOmega^{-1}$), 1.5 mm
  ($\tau_{\rm f}=0.003 \varOmega^{-1}$), 5 mm ($\tau_{\rm f}=0.01
  \varOmega^{-1}$) and 1.5 cm ($\tau_{\rm f}=0.03 \varOmega^{-1}$).
  Simulations start with a mean dust-to-gas ratio of $Z=0.01$, but gas is
  removed on a time-scale of 30 orbits (1 orbit = $2 \pi \varOmega^{-1}$),
  increasing the dust-to-gas ratio accordingly . While cm-sized particles
  concentrate in overdense filaments already at a modest increase in
  dust-to-gas ratio to $Z=0.015$, smaller particles require consecutive more
  gas removal to trigger clumping.}
  \label{f:carrera}
\end{figure*}

Computer simulations which follow the evolution of the streaming instability
into its non-linear regime show the emergence of axisymmetric filaments with
typical separations of 0.2 times the gas scale-height \citep{YangJohansen2014}
and local particle densities reaching several {\it thousand} times the gas
density \citep{BaiStone2010b,Johansen+etal2012}. These high densities trigger
the formation of large planetesimals (100--1000 km in diameter) by
gravitational fragmentation of the filaments \citep{Johansen+etal2007},
although planetesimal sizes decrease to approximately 100 km for a particle
column density comparable to that of the solar protoplanetary disk
\citep{Johansen+etal2012}. 

An important question concerning planetesimal formation through the streaming
instability is whether the process can operate for particles as small as
chondrules in the asteroid belt. In Figure \ref{f:carrera} we show numerical
experiments from \cite{Carrera+etal2015} on the streaming instability in
particles with sizes down to a fraction of a millimeter. The streaming
instability requires a threshold particle mass-loading $Z=\varSigma_{\rm
p}/\varSigma_{\rm g}$, where $\varSigma_{\rm p}$ and $\varSigma_{\rm g}$ are
the particle and gas column densities, to trigger the formation of overdense
filaments \citep{Johansen+etal2009b,BaiStone2010b}. The simulations in Figure
\ref{f:carrera} start at $Z=Z_0=0.01$, but the particle mass-loading is
continuously increased by removing the gas on a time-scale of 30 orbital
periods. This was done to identify how the critical value of $Z$ depends on the
particle size. The result is that overdense filaments form already at $Z=0.015$
for cm-sized particles, while large chondrules of mm sizes require $Z=0.04$ to
trigger filament formation. Chondrules smaller than mm do not form filaments
even at $Z=0.08$.

A lowered gas column density may thus be required to trigger concentration of
chondrule-sized particles by the streaming instability. It is possible that the
solar protoplanetary disk had a lower gas density than what is inferred from
the current mass of rock and ice in the planets (which multiplied by 100 gives
the MMSN), if the planet-forming regions of the nebula were fed by pebbles
drifting in from larger orbital distances \citep{Birnstiel+etal2012a}. In this
picture the growing planetesimals and planets are fed by drifting pebbles, so
that the current mass of the planets was achieved by the integrated capture
efficiency of the drifting solids; this allows for gas column densities lower
than in the Minimum Mass Solar Nebula to be consistent with the current masses
of planets in the Solar System. The gas will also be removed by accretion and
photoevaporation \citep{AlexanderArmitage2006}. The high mass-loading in the
gas could be obtained through pile-up by radial drift and release of refractory
grains near the ice line \citep{Sirono2011}.

Turbulence as weak as $\alpha\sim10^{-7}$ is necessary to allow the
sedimentation of chondrules (with ${\rm St}\sim10^{-3}$) into a thin mid-plane
layer with scale-height $H_{\rm p}=0.01 H_{\rm g}$ and $\rho_{\rm p} \approx
\rho_{\rm g}$ (see equation \ref{St}), the latter being a necessary density
criterion for activating particle pile up by streaming instabilities. Very low
levels of $\alpha$ are consistent with protoplanetary disk models where angular
momentum is transported by disk winds and the mid-plane remains laminar
\citep{BaiStone2013}, except for mild stirring by Kelvin-Helmholtz
\citep{YoudinShu2002} and streaming instabilities \citep{BaiStone2010b}.

Weak turbulence also facilitates the formation of dm-sized chondrule aggregates
\citep{Ormel+etal2008}, which would concentrate much more readily in the gas.
Stirring by hydrodynamical instabilities in the mid-plane, such as the vertical
shear instability \citep{Nelson+etal2013}, would preclude significant
sedimentation of chondrule-sized particles and affect the streaming
instability, as well as the formation of chondrule aggregates, negatively. An
alternative possibility is that the first asteroid seeds in fact did {\it not}
form from chondrules (or chondrule aggregates), but rather from larger icy
particles which would have been present in the asteroid formation region in
stages of the protoplanetary disk where the ice line was much closer to the
star \citep{MartinLivio2012,RosJohansen2013}. Chondrules could have been
incorporated into by later chondrule accretion (see Section \ref{s:layered}).

\section{\textbf{LAYERED ACCRETION}}
\label{s:layered}

The turbulent concentration model and the streaming instability, reviewed in
the previous sections, are the leading contenders for primary accretion of
chondrules into chondrites. However, none of the two are completely successful
in explaining the dominance of chondrules in chondrites: the turbulent
concentration models may not be able to concentrate sufficient amounts for
gravitational collapse, while the streaming instability relies on the formation
of chondrule aggregates and/or gas depletion and pile up of solid material from
the outer parts of the protoplanetary disk. In the layered accretion model the
chondrules are instead accreted onto the growing asteroids over millions of
years after the formation of the first asteroid seeds -- those first seeds
forming by direct coagulation from a population of 100-m-sized planetesimals as
envisioned in \cite{Weidenschilling2011} or by one or more of the particle
concentration mechanisms described in the previous sections.

\subsection{Chondrule accretion}

Chondrules are perfectly sized for drag-force-assisted accretion onto young
asteroids. The ubiquity of chondrules inside chondrites, and their large age
spread \citep{Connelly+etal2012}, indicates that planetesimals formed and
orbited within a sea of chondrules. Chondrules would have been swept past these
young asteroids with the sub-Keplerian gas. The gas is slightly
pressure-supported in the radial direction and hence moves slower than the
Keplerian speed by the positive amount $\Delta v$
\citep{Weidenschilling1977a,Nakagawa+etal1986}. The Bondi radius $R_{\rm B}=G
M/(\Delta v)^2$ marks the impact parameter for gravitational scattering of a
chondrule by an asteroid of mass $M$, with
\begin{equation}
  \frac{R_{\rm B}}{R} = 0.87 \left(
  \frac{R}{\rm 50\,km} \right)^2 \left( \frac{\Delta v}{\rm 53\,m/s}
  \right)^{-2} \left( \frac{\rho_\bullet}{\rm 3.5\,g/cm^3} \right) \, .
\end{equation}
Here we have normalised by $\Delta v=53$ m/s, the nominal value in the Minimum
Mass Solar Nebula model of \cite{Hayashi1981}, and used the chondrule density
$\rho_\bullet=3.5$ g/cm$^3$ as a reference value. Chondrules with friction time
comparable to the Bondi time-scale $t_{\rm B}=R_{\rm B}/\Delta v$ are accreted
by the asteroid
\citep{JohansenLacerda2010,OrmelKlahr2010,LambrechtsJohansen2012}. The
accretion radius $R_{\rm acc}$ can be calculated numerically as a function of
asteroid size and chondrule size by integrating the trajectory of a chondrule
moving with the sub-Keplerian gas flow past the asteroid. The accretion radius
peaks at $R_{\rm acc} \approx R_{\rm B}$ for $t_{\rm f}/t_{\rm B}$ in the range
from 0.5 to 10 \citep{LambrechtsJohansen2012}. Accretion at the full Bondi
radius happens for particle sizes
\begin{eqnarray}
  a &=& [0.008,0.16]\,{\rm mm}\, \left( \frac{R}{50\,{\rm km}} \right)^3 \left(
  \frac{\Delta v}{53\,{\rm m/s}} \right)^{-3} \nonumber \\ && \times
  \left( \frac{r}{2.5\,{\rm AU}} \right)^{-3} \left( \frac{\varSigma_{\rm
  g}}{\varSigma_{\rm MMSN}} \right) \, .
\end{eqnarray}
An asteroid of radius 50 km thus ``prefers'' to accrete chondrules of sizes
smaller than 0.1 mm, corresponding to the smallest chondrules found in
chondrites. At 100 km in radius the preferred chondrule size is closer to 0.2
mm, a 200 km radius body prefers mm-sized chondrules and larger bodies can only
grow efficiently if they can accrete chondrules of several mm or cm in
diameter. Carbonaceous chondrites accreted significant amounts of CAIs and
matrix together with their chondrules; \cite{Rubin2011} suggested that matrix
was accreted in the form of cm-sized porous aggregates with aerodynamical
friction time comparable to chondrules and CAIs.

Aerodynamical accretion of chondrules could explain the narrow range of
chondrule sizes found in the various classes of meteorites. The model predicts
that asteroids accrete increasingly larger chondrules as they grow. This
prediction may be at odds with the little variation in chondrule sizes found
within chondrite classes \citep[70\% of EH3 and CO3 chondrules have apparent
diameters within a factor of 2 of the mean apparent diameters in the group,
according to][]{Rubin2000}. The least metamorphosed LL chondrites nevertheless
do seem to host on the average larger chondrules \citep{NelsonRubin2002}. More
metamorphosed LL chondrites actually show a {\it lack} of small chondrules;
this could be due to the fact that the smallest chondrules disappeared from the
strongly heated central regions of the parent body.
\begin{figure}[!t]
  \begin{center}
    \includegraphics[width=\linewidth]{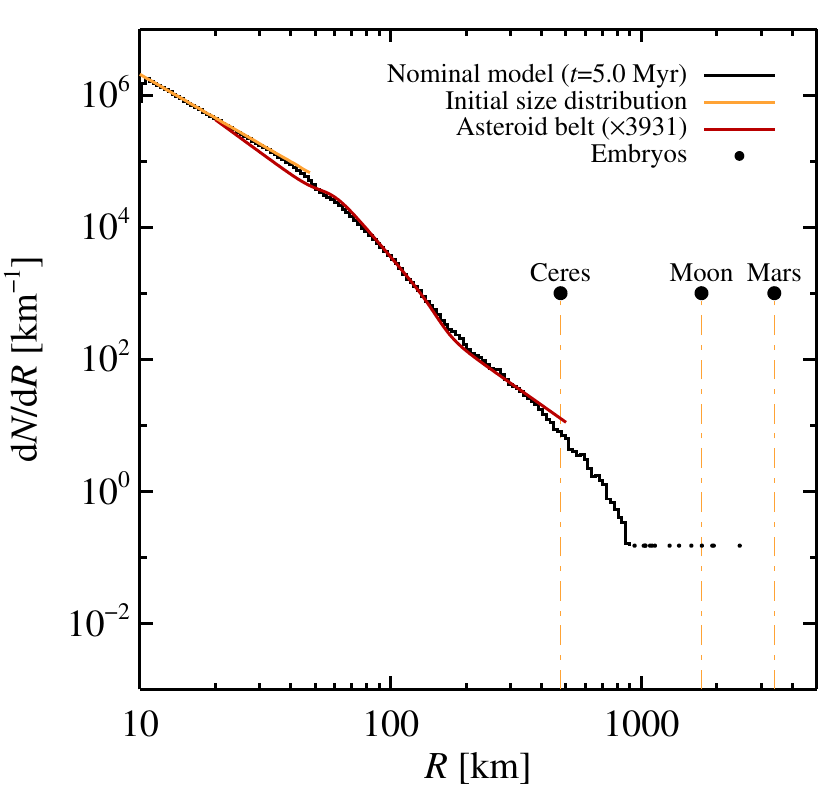}
  \end{center}
  \caption{The size distribution of asteroids and planetary embryos after
  accreting chondrules with sizes from 0.1 to 1.6 mm in diameter for 5 Myr. The
  original asteroid sizes had sizes between 10 and 50 km in radius (yellow
  line), here envisioned to form by the streaming instability in a population
  of dm-sized icy particles. The resulting size distribution of asteroids is in
  good agreement with the bump at 70 km in radius, the steep size distribution
  from 70 km to 200 km, and the shallower size distribution of larger asteroids
  whose chondrule accretion is slowed down by friction within their very large
  Bondi radius. Figure based on the results of \cite{Johansen+etal2015}.}
  \label{f:planetesimal_novariation}
\end{figure}

The accretion rate of chondrules (and other macroscopic particles) is
\begin{equation}
  \dot{M} = \pi f_{\rm B}^2 R_{\rm B}^2 \rho_{\rm p} \Delta v \, .
  \label{eq:mdot}
\end{equation}
Here $f_{\rm B}$ parameterises the actual accretion radius relative to the
Bondi radius and $\rho_{\rm p}$ is the chondrule density. Accretion of
chondrules is a run-away process, since $\dot{M} \propto R_{\rm B}^2 \propto
M^2$ if the optimal chondrule size is present (so that $f_{\rm B}=1$ in
equation \ref{eq:mdot}). The characteristic growth time-scale is
\begin{eqnarray}
  t_{\rm exp} &=& \frac{M}{\dot{M}} = 1.66\,{\rm Myr}\, \left(
  \frac{R}{50\,{\rm km}} \right)^{-3} \left( \frac{\Delta v}{53\,{\rm m/s}}
  \right)^{3} \nonumber \\
  && \hspace{-1.4cm} \times \left( \frac{r}{2.5\,{\rm AU}} \right)^{2.75} \left(
  \frac{\varSigma_{\rm g}}{\varSigma_{\rm MMSN}} \right)^{-1}
  \left( \frac{\rho_\bullet}{\rm 3.5\,g/cm^3} \right)^{-1} \, .
  \label{eq:texp}
\end{eqnarray}
We assumed here that the chondrules have sedimented to a thin mid-plane layer
of thickness 1\% relative to the gas scale-height. The strong dependence of the
accretion rate on the planetesimal mass will drive a steep differential size
distribution of a population of planetesimals accreting chondrules. This is
illustrated in Figure \ref{f:planetesimal_novariation} where asteroid seeds
with initial sizes from 10 to 50 km in radius have been exposed to chondrule
accretion over 5 Myr. The value of the turbulent viscosity is
$\alpha=2\times10^{-6}$. The resulting differential size distribution (which
manifests itself after around 3 Myr of chondrule accretion) shows a bump at 70
km in radius, a steep decline towards 200 km in radius, and finally a more slow
decline towards larger asteroids. The shallower decline is caused by the lack
of cm-sized particles needed to drive the continued run-away accretion of large
asteroids (this can be seen as a drop in $f_{\rm B}$ in equation
\ref{eq:mdot}). All the features in the size distribution in Figure
\ref{f:planetesimal_novariation} are in good agreement with features of the
observed size distribution of asteroids which are not explained well in
coagulation models \citep{Morbidelli+etal2009}. 

Layered accretion of chondrules can readily explain the large age spread of
individual chondrules inside chondrites \citep{Connelly+etal2012}, as well as
the remanent magnetisation of the Allende meteorite
\citep{Elkins-Tanton+etal2011}, imposed on the accreted chondrules from the
internal dynamo in the parent body's molten core. Efficient chondrule accretion
requires, as does the streaming instability discussed in the previous section,
sedimentation of chondrules to a thin mid-plane layer. The turbulent viscosity
of $\alpha=2\times10^{-6}$ in Figure \ref{f:planetesimal_novariation} is
nevertheless significantly larger than the $\alpha\sim10^{-7}$ needed to
sediment chondrules to a thin mid-plane layer of thickness 1\% of the gas scale
height; and indeed the time-scale to grow to the current asteroid population is
3 Myr in the simulation shown in Figure 4, about twice as long as in equation
(\ref{eq:texp}), but in good agreement with the ages of the youngest
chondrules.

\section{\textbf{OPEN QUESTIONS}}
\label{s:questions}

The formation of asteroids is a complex problem which will only be solved
through a collective effort from astronomers, planetary scientists and
cosmochemists. Although many details of asteroid formation are still not
understood, we hope to have convinced the reader that new insights have been
achieved in many areas in the past years. Here we highlight ten areas of open
questions in which we believe that major progress will be made in the next
decade:

1) {\it Short-lived radionuclides}. What is the origin of the short-lived
radioactive elements which melted the differentiated parent bodies? Was
$^{26}$Al heterogeneous in the solar protoplanetary disk? How did the young
Solar System become polluted in $^{26}$Al without receiving large amounts of
$^{60}$Fe, an element that is copiously produced in supernovae?

2) {\it Maintaining free-floating chondrules and CAIs}. How is it possible to
preserve chondrules and CAIs for millions of years in the disk before storing
them in a chondritic body, without mixing them too much to erase chondrule
classes and chondrule-matrix complementarity? What are we missing that makes
this issue so paradoxical?

3) {\it Chondrules versus matrix}. Why do carbonaceous chondrites contain large
amounts of matrix while ordinary chondrites contain very little matrix? Did the
matrix enter the chondrites as (potentially icy) ``matrix lumps'' or on
fine-grained rims attached to chondrules and other macroscopic particles?

4) {\it Initial asteroid sizes.} What is the origin of the steep differential
size distribution of asteroids beyond the knee at 100 km?  Did asteroids form
small as in the coagulation picture, medium-sized as in the layered accretion
model or large as in some turbulent concentration models? Why do Kuiper belt
objects, which formed under very different conditions in temperature and
density, display a similar size distribution as asteroids?

5) {\it The origin of asteroid classes}. How is the radial gradient of asteroid
composition produced and retained in the presence of considerable
pre-accretionary turbulent mixing and post-accretionary dynamical mixing?  Is
asteroid formation a continuous process which happens throughout the life-time
of the protoplanetary disk? What do the different chondrite groups mean in
terms of formation location and time?

6) {\it Dry and wet chondrites}. Why do we have dry chondrites (enstatite,
ordinary)? If chondrites form at 2-4 Myr after CAIs, then the snowline should
have been well inside the inner edge of the asteroid belt. Are there overlooked
heating sources which could keep the ice line at 3 AU throughout the life-time
of the protoplanetary disk? Or did the asteroid classes form at totally
different places only to be transported to their current orbits later?

7) {\it Internal structure of asteroids}. Does the chondrite and asteroid
family evidence suggest that the primary asteroids -- before internal
heating -- are homogeneous, roughly 100-km-diameter bodies composed of a
physically, chemically, and isotopically homogeneous mix of chondrule-sized
components? Or is internal heterogeneity, as may be the case for the Allende
parent body, prevalent?

8) {\it Turbulent concentration of chondrules}. Under what nebula conditions
can vortex tubes over a range of nebula scales concentrate enough chondrules
into volumes that are gravitationally bound, at a high enough rate to produce
the primordial asteroids and meteorite parent bodies directly? What other roles
could turbulent concentration play in planetesimal formation given that the
optimally concentrated particle is chondrule-sized under nominal values of the
turbulent viscosity?

9) {\it Streaming instability with chondrules}. Will the conditions for
streaming instabilities to concentrate chondrule-sized particles, i.e.\ gas
depletion and/or particle pile up, be fulfilled in the protoplanetary disk? How
does an overdense filament of chondrule-sized particles collapse under
self-gravity given the strong support by gas pressure?

10) {\it Layered accretion}. What is the origin of the apparent scarcity of
heterogeneous asteroid families, given that asteroids orbiting within an ocean
of chondrules should accrete these prodigiously? What is the thermal evolution
of early-formed asteroid seeds that continue to accrete chondrules over
millions of years?

{\bf Acknowledgments.} AJ was supported by the Swedish Research Council (grant
2010-3710), the European Research Council under ERC Starting Grant agreement
278675-PEBBLE2PLANET and by the Knut and Alice Wallenberg Foundation. He would
like to thank Benjamin Weiss for stimulating discussions on layered accretion.
EJ wishes to remember his colleague and friend Guillaume Barlet (1985-2014),
who as a short-lived radionuclide theorist and chondrule/refractory inclusion
specialist would certainly have contributed to the new paradigms discussed
herein, true to his attachment to interdisciplinary interactions, but left us
far too early. JC thanks Chris Ormel for a careful reading, and Ed Scott, Alan
Rubin, Noriko Kita, and Gerry Wasserburg for helpful comments and references.
We would like to thank Alan Rubin and an additional anonymous referee for
insightful referee reports.

\bibliographystyle{ppvi_lim3_with_title.bst}
\bibliography{bibliography}

\end{document}